\documentstyle[12pt]{article}
\begin{document}

\def\rg{\rangle }
\def\lg{\langle }
\def\ra{\rightarrow }
\def\AS{{\cal A}}
\def\BS{{\cal B}}
\def\HS{{\cal H}}

\title{Semiclassical Fourier Transform for Quantum Computation}

\author{Robert B. Griffiths\thanks{Electronic mail: rgrif+@cmu.edu}\ \
and Chi-Sheng Niu\thanks{Electronic mail: cn28+@andrew.cmu.edu}\\
Department of Physics\\
Carnegie Mellon University\\
Pittsburgh, PA 15213, U.S.A.}

\date{Version of 3 Nov. 1995}
\maketitle

\begin{abstract}
	Shor's algorithms for factorization and discrete logarithms on a
quantum computer employ Fourier transforms preceding a final measurement.  It
is shown that such a Fourier transform can be carried out in a semi-classical
way in which a ``classical'' (macroscopic) signal resulting from the
measurement of one bit (embodied in a two-state quantum system) is employed to
determine the type of measurement carried out on the next bit, and so forth.
In this way the two-bit gates in the Fourier transform can all be replaced by a
smaller number of one-bit gates controlled by classical signals.  Success in
simplifying the Fourier transform suggests that it may be worthwhile looking
for other ways of using semi-classical methods in quantum computing.
\end{abstract}


	Recently Shor \cite{sh94,sh95} has shown that a quantum computer
\cite{dt8589}, if it could be built, would be capable of solving certain
problems, such as factoring long numbers, much more rapidly than is possible
using currently available algorithms on a conventional computer. This has
stimulated a lot of interest in the subject \cite{bn95,dv95b,lla95}, and
various
proposals have been made for actually constructing such a computer
\cite{ll93,bd95,sw95,cz95,cy95}.  The basic idea is that bits representing
numbers can be embodied in two-state quantum systems, for example, in the spin
degree of freedom of a spin half particle, and the computation proceeds by
manipulating these bits using appropriate gates. It turns out that quantum
computations can be carried out using circuits employing one-bit gates, which
produce a unitary transformation on the two-dimensional Hilbert space
representing a single bit, together with two-bit gates producing appropriate
unitary transformations on a four-dimensional Hilbert space
\cite{dv95a,dbe95,llb95,bal95}.

	One-bit gates should be much easier to construct than two-bit gates,
since, for example, an arbitrary unitary transformation on the spin degree of
freedom of a spin half particle can be produced by subjecting it to a suitable
time-dependent macroscopic magnetic field.  On the other hand, a two-bit gate
requires that one of the bits influence the other in a non-trivial way, and
this without leaving any record in the environment, since the computer utilizes
coherent quantum states.

	In this letter we shall show how the quantum Fourier transforms which
in Shor's algorithms immediately precede a final measurement can be carried out
in a semi-classical fashion which requires no two-bit gates.  The trick is to
measure a particular bit and then use the result to produce a classical signal
which controls a one-bit transformation carried out on the next bit just before
it is measured, and so forth.  It is of interest for at least three reasons.
First, it represents a completely new (so far as we know) technique for quantum
computation, in which the results of certain measurements, converted into
``classical'' signals (imagine a pulse of several volts traveling down a
coaxial cable), can be used to influence a later step in the computation
\cite{tport}.  Second, computing the Fourier transform is considerably
simplified in the sense that it requires no two-bit gates. Third, a simple way
of seeing why the semi-classical method actually works is to adopt a point of
view---in technical terms, a particular family of consistent histories---in
which the results of a measurement are used to infer the state of a quantum
system before the measurement was carried out.  This perspective may prove
useful in thinking about other issues in quantum computing, quantum optics, and
quantum effects in atomic physics.

	Since Shor has described his algorithms in considerable detail in his
publications \cite{sh94,sh95}, we shall not discuss them here.  It will be
sufficient to note that after a certain number of steps of the quantum
computation have been carried out, the relevant quantum state $|\psi\rg$ is a
coherent superposition \cite{n01}
of different states $|a\rg$ labeled by an integer $a$ between 0 and $q-1$,
where $q=2^{s+1}$.  The state $|\psi\rg$ is then subjected to a unitary
transformation $F$, a sort of discrete Fourier transform, which carries
each basis state $|a\rg$ of the $q$-dimensional Hilbert space into
\begin{equation}
 F |a\rg = {1\over\sqrt q} \sum_{c=0} ^{q-1} e^{2\pi iac/q} |c\rg.
\label{e.1}
\end{equation}
This is followed by a measurement of the integer $c$, that is, a
measurement of each of its $s+1$ bits.

	A set of gates which carries out this Fourier transform \cite{r1} is
shown in Fig.~1 for $s=3$. The bits to be transformed enter from the left.  One
can imagine that they are spin 1/2 particles, with the results of the preceding
computation embodied (as a coherent superposition) in their collective spin
degrees of freedom.  These particles then move through a series of gates,
indicated by circles, and eventually arrive at measuring devices, shown as
squares, where a particular component of spin, say $S_z$, is measured by a
Stern-Gerlach device, with $S_z=1/2$ in units of $\hbar$ interpreted as the bit
$|1\rg$, and $S_z=-1/2$ as the bit $|0\rg$.

	If the binary representation of the number $a$ is
\begin{equation}
  a=\sum_{j=0}^s a_j 2^j,
\label{e.2}
\end{equation}
with each $a_j$ zero or one, the state $|a\rg$ can be conveniently written as a
tensor product
\begin{equation}
 |a\rg = |a_s a_{s-1}\ldots a_0\rg =
|a_s\rg_s\otimes\cdots |a_1\rg_1\otimes |a_0\rg_0.
\label{e.3}
\end{equation}
Using the same notation for $|c\rg$, we can rewrite (\ref{e.1})
in the form
\begin{equation}
 F |a\rg = \prod_{j=0}^s \otimes |p(\phi_j)\rg_j,
\label{e.4}
\end{equation}
where the state
\begin{equation}
 |p(\phi)\rg = \bigl( |0\rg + e^{2\pi i\phi} |1\rg\bigr)/\sqrt 2
\label{e.5}
\end{equation}
will be said to have a {\it phase} $\phi$ between 0 and 1.
The phases $\phi_j$ in (\ref{e.4}) take the values:
\begin{equation}
 \phi_j = \sum_{k=0}^{s-j} a_k 2^{j+k-s-1}.
\label{e.6}
\end{equation}
In terms of our picture of a spin half particle, $|p(\phi)\rg$ represents a
spin component of $1/2$ in a direction in the $x,y$ plane determined by
$\phi$.

	The one-bit gates in the top row of Fig.~1 transform the bit entering
on the left to one leaving on the right through:
\begin{equation}
	\begin{array}{r@{}l}
 |0\rg \ra \bigl(|0\rg + |1\rg\bigr)/\sqrt 2 &\;= |p(0)\rg,\\
 |1\rg \ra \bigl(|0\rg - |1\rg\bigr)/\sqrt 2 &\;= |p(1/2)\rg.\\
	\end{array}
\label{e.7}
\end{equation}
(In a spin picture, the spin is rotated by $90^\circ$ about the $y$ axis.)
The two bit gate labeled with an integer $m$ converts the bits entering on the
left into those leaving on the right according to the scheme:
\begin{equation}
	\begin{array}{r@{}l}
 |00\rg \ra |00\rg,\quad &|01\rg \ra |01\rg,\\
 |10\rg \ra |10\rg,\quad &|11\rg \ra \exp[2\pi i/2^m]|11\rg.
	\end{array}
\label{e.8}
\end{equation}
Here we use the convention that the left element in each ket $|\rg$ in
(\ref{e.8}) is the bit which in Fig.~1 enters the two-bit gate at a point
marked by a black dot and also leaves at a black dot, while the right element
in each ket denotes the bit which enters and leaves the gate at a point labeled
by a circle.

	It is helpful to think of (\ref{e.8}) as a transformation in which one
bit acts as a ``control'' which enters and leaves the gate as a zero or one,
while the other ``target'' bit enters as $|p(\phi)\rg$ and undergoes a phase
shift, so that it leaves as $|p(\phi')\rg$, where $\phi'=\phi+2^{-m}$.  From
this perspective it is easy to see how the network in Fig.~1 produces the
desired result.  Suppose that the bit $a_3$ enters the left-most one-bit gate
in Fig.~1 in the state $|0\rg$.  It emerges at the point $B$ in the state
$|p(0)\rg$.  As it passes downwards through the successive two bit gates its
phase is shifted by the bits $a_2$, $a_1$, and $a_0$, acting as control bits,
by an amount
\begin{equation}
	\Delta\phi=a_0/16 + a_1/8 + a_2/4.
\label{e.9}
\end{equation}
If, on the other hand, $a_3$ arrives as $|1\rg$, the one-bit gate converts it
to $|p(1/2)\rg$, so that the bit which emerges as $c_0$ has a phase $\Delta\phi
+ a_3/2$, in agreement with (\ref{e.6}).  The same sort of analysis can be
applied to the rest of the circuit.

	However, one can equally well regard the bits entering and leaving the
gates through the black dots in Fig.~1 as the control bits, and it is this
point of view which is useful for constructing the semiclassical Fourier
transform.  Suppose that the final measurement reveals that the bit $c_0$ is 1,
corresponding to $|1\rg$. Then, since a control bit enters and leaves the
two-bit gates unchanged, we conclude that this bit was also in the state
$|1\rg$ at point $B$, just after emerging from the first one-bit gate.
Similarly, if the measurement yields $c_0=0$, we conclude that the bit was in
the state $|0\rg$ at the point $B$.  Hence the circuit would work equally well
if the measurement on this bit were actually carried out at the point $B$ in
Fig.~1, were it not for the fact that this bit is also needed in order to
influence $a_2$, $a_1$, and $a_0$, now regarded as target bits, as they pass
through the two-bit gates controlled by $c_0$.  However, if $c_0$ is measured
at $B$ and the result is converted to a classical signal, this signal
can be used
to determine the action of a corresponding set of one-bit gates acting upon
$a_2$, $a_1$, and $a_0$.

	Applying this type of analysis to the other parts of Fig.~1, one is led
to the arrangement shown in Fig.~2, where all the two bit gates in Fig.~1 have
been eliminated, and their work taken over by one-bit gates
controlled by classical signals (double lines) and  followed by
measurements.  Each of the boxes in Fig.~2 performs the following
operations.  The incoming bit is first subjected to a unitary transformation,
equivalent to a phase shift followed by (\ref{e.7}):
\begin{equation}
	\begin{array}{r@{}l}
	|0\rg &\;\ra \bigl( |0\rg + |1\rg\bigr) /\sqrt 2, \\
	|1\rg &\;\ra e^{2\pi i\phi}\bigl( |0\rg - |1\rg\bigr) /\sqrt 2, \\
	\end{array}
\label{e.10}
\end{equation}
where $\phi$ is the phase transmitted as a classical signal from the previous
box.  The bit is then measured to yield the result $c=0$ or $1$.  The outputs
are two classical signals represented by double lines: one is the result of the
measurement, and the other represents a phase
\begin{equation}
  \phi' = \phi/2 + c/4
\label{e.11}
\end{equation}
which is sent to the next box. The very first box uses $\phi=0$ in
(\ref{e.10}).

	Readers unfamiliar with the rules which allow one to use the results of
a measurement to infer the state of a quantum system prior to the measurement
are referred to the relevant literature \cite{gr84,gr86,om92,gr93,om94}.  The
basic idea is that one can add to a quantum history, consisting of an initial
state and the result of a measurement, a projector, for example
\begin{equation}
	P=|c_0=1\rg\lg c_0=1|,
\label{e.12}
\end{equation}
representing property of the quantum system at a time just prior to the
measurement.  The conditional probability of $P$ is one, given that the
measurement yields $c_0=1$.  But in the situation shown in Fig.~1, $P$ commutes
with the unitary transformations corresponding to the three 2-bit gates which
precede the measurement of $c_0$, and therefore one can ``push'' the
corresponding property backwards in time to the point $B$ in Fig.~1, and again
conclude that it occurred with probability one.  Anyone who feels uncomfortable
with this method of reasoning can, of course, use traditional techniques to
verify that the arrangement in Fig.~2 will yield the same probability to
observe a number $c$ as that in Fig.~1. (Note that it is necessary to check
this for an initial state $|\psi\rg$ which is an arbitrary linear combination
of the different $|a\rg$ states.)

	The scheme in Fig.~2 should be quite a bit simpler to realize than that
in Fig.~1, because it only requires one-bit operations controlled by classical
signals rather than the more difficult and more numerous two-bit operations
indicated in Fig.~1.  However, the need to carry out a measurement on one bit
before beginning to measure another could be a disadvantage if the physical
elements representing the bits are short lived or decohere rapidly on the time
scale required to carry out a measurement and convert it into a classical
signal. If this turns out to be a problem, a possible remedy might be to
arrange the earlier part of the quantum computation in such a way that the more
significant bits of $a$ are produced earlier than the less significant bits.

	An important question is whether other parts of Shor's algorithms (or
other applications of quantum computing) can make use of similar semi-classical
operations.  We have no specific proposals, but we think the idea is worth
further study.  There are bits other than those entering the final Fourier
transform which are produced elsewhere in the computation, and it is
conceivable that measuring some of these could be used in modify later steps in
the calculation, or perhaps for the non-trivial task of correcting errors.  In
connection with the latter, see \cite{sh95b}.

	Finally, we note that adopting alternative perspectives or points of
view about what is going on in a quantum computation can yield useful insights.
The traditional perspective, represented in almost all work on quantum
computing up till now, in which the ``wave function of the computer'' develops
unitarily in time until a measurement is made, has demonstrated its value
through the work of various people who have brought quantum computation to
its current state of development.  In this letter we have shown that an
alternative point of view, in which the results of measurements are traced
backwards in time, can also be valuable. The general principles of quantum
mechanics allow for a variety of viewpoints or, to use the technical term,
consistent families, and recent developments in the foundations of quantum
theory \cite{gr94} permit one to make effective use of these without becoming
entangled in paradoxes, mysterious long-range influences, and the like.


\section*{Acknowledgements}

	  We are grateful to Dr.~D.~DiVincenzo for some helpful comments on the
manuscript. Financial support for this research has been provided by the
National Science Foundation through grant PHY-9220726.




\section*{Figure Captions}

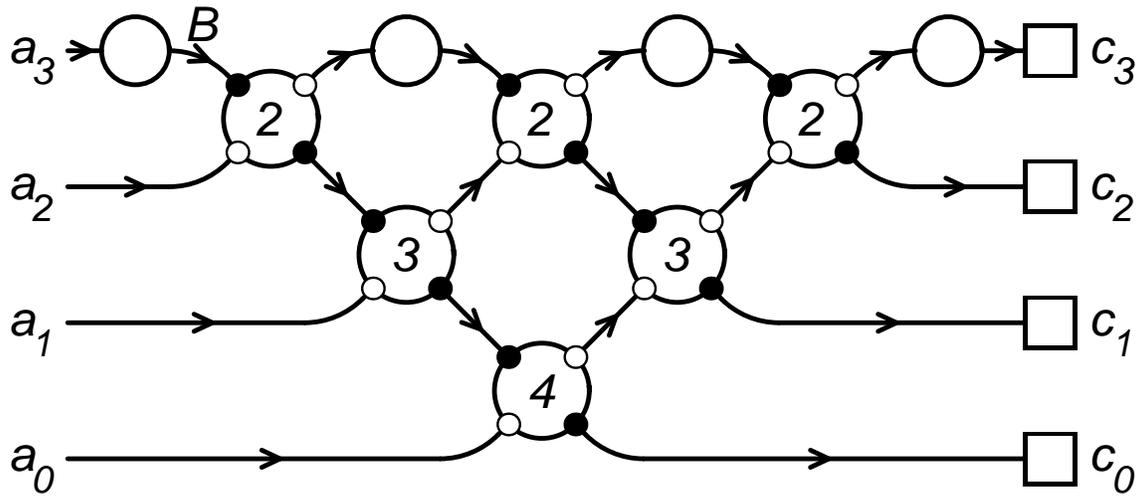
\begin{figure}[h]
\caption{Arrangement of quantum gates to carry out the discrete Fourier
transform (\ref{e.4}) on numbers consisting of four bits.}
\label{fig1}
\end{figure}

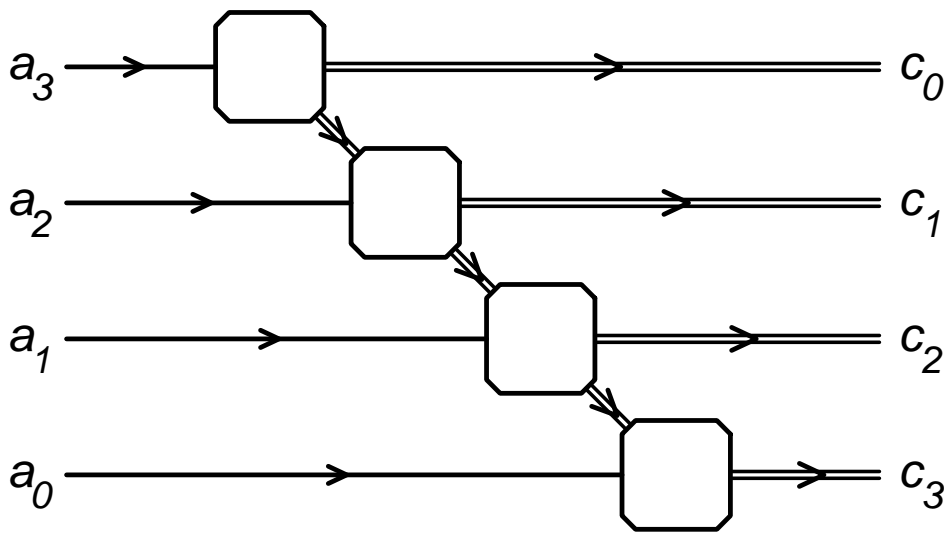
\begin{figure}[h]
\caption{Semiclassical procedure which produces the same result as the circuit
in Fig.~1. The double lines represent classical signals.}
\label{fig2}
\end{figure}


\begin{thebibliography}{11}
\bibitem{n01}
Or, if one prefers, a  reduced density matrix;
this makes no difference for the following discussion.
\end{thebibliography}
\end{document}